\documentclass[amsmath,amssymb,preprint]{revtex4-1}
\usepackage{graphicx}  % needed for figures
\usepackage{dcolumn}   % needed for some tables
\usepackage{bm}   % for math

\usepackage[utf8]{inputenc}
\begin{document}
\title{Testing Case Number of Coronavirus Disease 2019 in China with Newcomb-Benford Law}
%\author{Junyi Zhang (张骏祎)}
\author{Junyi Zhang}
\email{junyiz@princeton.edu.}
\affiliation{Department of Physics, Princeton University, Princeton 08544, New Jersey, USA.}

\date{\today}% It is always \today, today,

\begin{abstract}
The coronavirus disease 2019 bursted out about two months ago in Wuhan has caused the death of more than a thousand people.  China is fighting hard against the epidemics with the helps from all over the world.  On the other hand, there appear to be doubts on the reported case number.  
In this article, we propose a test of the reported case number of coronavirus disease 2019 in China with Newcomb-Benford law.
We find a $p$-value of $92.8\%$ in favour that the cumulative case numbers abide by the Newcomb-Benford law.  
Even though the reported case number can be lower than the real number of affected people due to various reasons, this test does not seem to indicate the detection of frauds.
\end{abstract}

\maketitle

\section{\label{sec:Intro}Introduction}
The coronavirus disease 2019 (COVID19), first observed in mid December 2019 in Wuhan China, has hitherto (as reported by Feb. 12, 2020) caused $45,171$ confirmed cases among which $44730$ are in China and $1115$ deaths among which $1$ is outside of China~\cite{WHOFeb12}.
World Health Organization declared the outbreak to be a public health emergency of international concern~\cite{WHOJan31}.

China is facing the huge challenges of providing timely health care for the patients and preventing further spread of COVID19.  Although many countries and individuals provide various supports to China for fighting against the virus, hospitals, especially those in small counties outside Wuhan, are still short of biomedical equipment and other necessary  supplies.  A severe impact on Chinese economy is also expected due to the outbreak of the disease.  If the spread of COVID19 was not controlled effectively due to its high basic reproduction number $R_0$ and long incubation period, there might be a worse impact world wide.

At the same time, the case number reported by the Chinese government has been questioned  in worry of the government intentionally hiding the real situation.  In this article, we provide a quantitative analysis on the data of the COVID19 reported.  A test of frauds with Newcomb-Benford law has been implemented to examine the data of the reported case numbers  of COVID19 in China.  We find a $p$-value of $92.8\%$ in favour that the cumulative case numbers of COVID19 in China abide by the Newcomb-Benford law.  Therefore this test does not seem to indicate the detection of frauds.
Nevertheless, it is not conclusive the reported case number precisely reflects the current situations.
The details of the hypothesis and testing will be shown in Sec.~\ref{sec:Test}.  Further discussion is presented in Sec.~\ref{sec:Discussion}.

\section{\label{sec:Test}Hypothesis and Testing}
The skewed distribution of the significant digits was discovered by Newcomb~\cite{Newcomb1881} and latter advanced by Benford~\cite{Benford1938}.
They found that the frequencies of the first significant digits $d$ of the numbers in some statistics follow the distribution
\begin{equation}\label{eq:NBLawDistribution}
P_{NB}(d) = \log_{10} \left(1 + \frac{1}{d}\right),
\end{equation}
which is now often referred to as Newcomb-Benford Law (NBL).
One may refer to Ref.~\citenum{Miller2015} for historical developments and and applications of NBL. 
Not all the distributions obey the NBL.  There are ones known to obey NBL exactly, e.g., Fibonacci numbers, factorials, powers of 2 and exponential growth process~\cite{Duncan1967,Sarkar1973,Washington1981,Miller2015},  while others known not to obey NBL e.g., square roots and reciprocals~\cite{Raimi1976}
An interesting application is to detect the frauds in financial activities even though no general theory proved NBL should hold~\cite{Miller2015,CG2007,CBCMP2019}.  

\begin{figure}[htbp]
\begin{center}
\includegraphics[scale = 0.7]{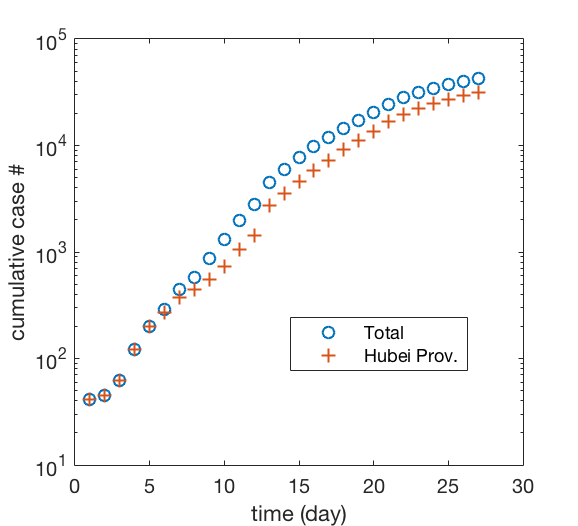}
\caption{COVID19 Cumulative Case Number.}
\label{fig:TotoCumulativeNumber}
\end{center}
\end{figure}

We examine the cumulative case numbers of COVID19 of $31$ province-level divisions from 15 Jan. 2020 to 10 Feb. 2020.  The data used were from wikipedia, ``{\sl Timeline of the 2019–20 Wuhan coronavirus outbreak in February 2020}''.  The cumulative case numbers were obtained from ``{\sl Tab. New Confirmed Cases of Coronavirus in Mainland China by Province-Level Divisions}''~\cite{WikiCOVID19Feb12} by adding the numbers according to the time order.
Fig.~\ref{fig:TotoCumulativeNumber} shows the total cumulative case number and the cumulative case number in Hubei province versus time  from 15 Jan. 2020 to 10 Feb. 2020 in a semilog plot.

For the first $15$ days, the total cumulative case number increases exponentially, which is not surprising according to  simple models of epidemics.
The slop of the more recent total cumulative case number in the semilog plot however decreases.   This transition around 30 Jan. seems to agree with the incubation period counting from 20-23 Jan. when several districts in China declared highest level emergency status for the epidemics, particularly when Wuhan was closed.
However, it is not conclusive that COVID19 has bee controlled and the total cumulative case number tends to saturate.  It is also possible that the total cumulative case number is still increasing exponentially but with a smaller rate.  The decreasing of the rate may be due to the emergent isolation strategy and better protections (everyone wearing face masks).  Nevertheless due to the limit medical resources, many suspected cases remained to be confirmed and people with extraordinary long incubation period are still potential sources for the further expansion.  Convection of the population after the Chinese New Year might worsen the situation and cause another explosion.

To the best knowledge of the author there are no general theories prove that the statistics of the epidemics like COVID19 should obey the NBL.  It will be interesting to test if the any statistics follows NBL.  It naturally has the potential applications as to detect frauds in the statistics of the epidemics.
A reasonable expectation might be the cumulative case numbers that follows the NBL, as it seems to grow exponentially particularly at the beginning stage.  
We count the frequency of the first significant of COVID19 according to province-level divisions.  Although the first few cases in each division might have contaminate the virus from Hubei Province, the spread afterwords within the division may be considered independent development but following the same propagation law of COVID19.  The inter-province convection of the  population was suppressed significantly due to the Chinese New Year vacation and isolation policy after 24 Jan.

We propose the null hypothesis ($H_0$): the cumulative case numbers of COVID19 obey the NBL.
\begin{figure}[htbp]
\begin{center}
\includegraphics[scale = 0.8]{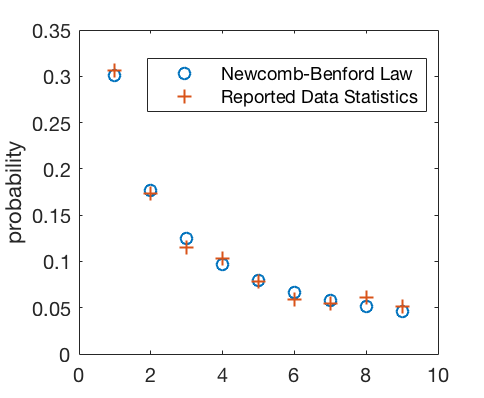}
\caption{Newcomb-Benford Law.}
\label{fig:NBL}
\end{center}
\end{figure}
Fig.~\ref{fig:NBL} shows both the theoretical distribution of the NBL according to Eq.~\ref{eq:NBLawDistribution} (with blue circles) and the empirical distribution of the first significant figure of the reported cumulative case numbers (with red cross).  To the first sight, they seem to agree quite well.

To quantify the agreement  between the empirical distribution and the hypothetical distribution, we may consider the more general hypothesis ($H'_0$) as follows.
Suppose the  cumulative case numbers of COVID19 follows the distribution $\Psi$ in the absence of fraud but $\Phi$ otherwise and the probability of the frauds is measured by $\tau$.
Then the hypothetical distribution $\Pi$ should be interpolated by the distribution with and without frauds~\cite{Miller2015}, i.e.,
\begin{equation}\label{eq:FraudNBLawDistribution}
\Pi = (1-\tau)\Psi + \tau \Phi.
\end{equation}
Then the hypothesis $H_0$ is nothing but the special case of $H'_0$ with $\tau =0$, i.e., in the absence of the frauds, the cumulative case numbers of COVID19 obey the NBL.

One of the simplest statistics to test the agreement of $\Pi$ and $P_{NB}$ can be chosen as
\begin{equation}\label{eq:Chi2Test}
V = \sum_d  \frac{\Big(N(d) - N_{tot} P_{NB}(d) \Big)^2}{N_{tot} P_{NB}(d) },
\end{equation}
where $N(d)$ is the counting of figure $d$ as the first significant digit among $N_{tot}$ cumulative case numbers. 
When $N_{tot} \rightarrow \infty$, $V$ is subjected to the $\chi_\nu^2$ distribution, where $\nu = 9 \times 10^{k-1}-1$~\cite{CBCMP2019}, $k=1$.
 
It easy to calculate $V$ directly according to Eq.~\ref{eq:Chi2Test}, we find $V = 3.10$ and the corresponding $p$-value is $92.8\%$.
Therefore, we conclude that the test is in favour of the null hypothesis $H_0$ that the cumulative case numbers of COVID19 obey the NBL.
If we assume that it is hard to fabricate data closely following the NBL~\cite{CBCMP2019}, we may conclude we did not detect frauds according to this test with the NBL.

\section{\label{sec:Discussion}Discussion}
In our hypothesis and testing, we showed that NBL holds for the cumulative case number of COVID19, nevertheless, same as in the case of financial activities, there are no general theory predicting the NBL~\cite{Miller2015,CG2007,CBCMP2019}.  For checking the agreement with the NBL, similar tests may be applied to other epidemics like the flus (HxNy), SARS {\sl etc.}  It deserve to notice that, COVID19 seems still in its expanding stage with exponential scaling, and we know the exponential grow process satisfies the NBL, while it still remains to be tested in the NBL is valid when epidemics are no longer in the initial exponentially growing stage.

We have in total $628$ data point for the statistics.  This is not a big sample set.  Simple $\chi^2$-test may be subjected to large fluctuations.  More advanced tests, like the Kolmogorov-Smirnov test or the Kuiper test may be used instead.

It seems quite reasonable to accept the null hypothesis $H_0$ that the cumulative case numbers of COVID19 obey the NBL. Taking this as an assumption we may apply the NBL to detect frauds in epidemic statistics.  Further more, we do not think frauds are detected  from  the cumulative case numbers of COVID19. 
Nevertheless, it does not mean that the reported cumulative case numbers precisely reflect the current situation exactly, particularly in Hubei Province.  Due to the lack of medical equipment and resources, many suspected cases remain to be confirmed; some people showing syndromes may not even be counted as suspected cases yet.  Therefore the statistical data can be biased.  Unfortunately, two exponential grow process differing by a multiplicative constant, both abide by the NBL.  This test with the NBL cannot tell if there are systematic fraction of patients not being included properly.

\section{Final Remarks}
The Earth is our common home of human beings and other creatures in nature.  Our technologies of transportation has advanced unprecedentedly over the last centuries, it also brings more and more possibilities of quick and wide spread of the epidemics.  We should stand together, help each other facing the crisis of the epidemics no matter of race, gender or religion.  Spring cannot be far behind.

\end{document}